\documentclass[%
pre%
,twocolumn
,showpacs
,preprintnumbers
]{revtex4}
\usepackage{amsmath} 
\usepackage{bm} 

\begin{document}



%
%

\title{Statistical Theory for Incoherent Light
Propagation in Nonlinear Media}


\author{B. Hall}
\email{bjorn.hall@elmagn.chalmers.se}
\author{M. Lisak}
\author{D. Anderson}
\affiliation{Dept.\ of Electromagnetics, Chalmers University of Technology,
SE-412 96, G\"oteborg, Sweden}

\author{R. Fedele}
\affiliation{Dipartimento di Scienze Fisiche, Universita ´Federico
II di Napoli and INFN Sezione di Napoli, Complesso Universitario
di MS Angelo, Via Cintia, I-80 126 Napoli, Italy }

\author{V. E. Semenov}
\affiliation{Institute of Applied Physics, Russian Academy of
Sciences, 46 Ulyanov Street, Nizhny Novgorod 603 600, Russia}

\begin{abstract}
A novel, statistical approach based on the Wigner transform is proposed for the
description of partially incoherent optical wave dynamics in nonlinear media.
An evolution equation for the Wigner transform is derived from a nonlinear
Schr\"odinger equation with arbitrary nonlinearity. It is shown that random
phase fluctuations of an incoherent plane wave lead to a Landau-like damping
effect, which can stabilize the modulational instability. In the limit of the
geometrical optics approximation, incoherent, localized, and stationary
wave-fields are shown to exist for a wide class of nonlinear media.
\end{abstract}


\maketitle

%
%

It is well known that the propagation of electromagnetic waves and
beams in dispersive nonlinear media is associated with the
phenomena of modulational instability and the formation of
stationary wave structures, i.e., temporal and spatial envelope
solitons, cf.\ \cite{Karpman}. The conventional treatment of these
phenomena considers coherent wave structures. However, a recent
series of experimental \cite{Mitchell} and theoretical papers
\cite{Pasmanik, Christodoulides, Hasegawa, Christodoulides2} has
demonstrated the existence of incoherent solitons. In particular,
spatially incoherent optical solitons have attracted much
attention. Until now, three approaches have been developed to
describe partially incoherent wave propagation in nonlinear media:
(i) the propagation equation for the mutual coherence function
\cite{Pasmanik}, (ii) the coherent density method
\cite{Christodoulides}, and (iii) the self-consistent multimode
theory \cite{Hasegawa}. In fact, it has been shown in Ref.\
\cite{Christodoulides2} that these three approaches are formally
equivalent.

In the present article, we propose a novel, general, statistical
theory for describing the dynamics of partially incoherent optical
waves and beams in dispersive and nonlinear media. The approach is
based on the Wigner transform method \cite{Wigner}, which was
introduced in statistical quantum mechanics to describe the
dynamics of the quantum state of a system in the classical space
language. A similar approach has been successfully applied in
plasma physics in connection with the theory of weak plasma
turbulence \cite{Venedov}, as well as with the description of
electromagnetic wave propagation in a nonstationary, inhomogenous,
and relativistic plasma \cite{Tsinsadze}. Recently, the Wigner
function method has been used to analyze the longitudinal dynamics
of charged-particle beams in accelerators \cite{Anderson}, and to
study the dynamics of Bose-Einstein condensates in the presence of
a chaotic external potential \cite{Gardiner}.

The scope and outline of this article is as follows. Starting from
the nonlinear Schr\"odinger equation describing the evolution of
the slowly varying wave amplitude in a dispersive medium with an
arbitrary nonlinear response, we derive the Wigner-Moyal equation
for the Wigner transform including the Klimontovich statistical
average. This equation reduces, in the geometrical optics
approximation, to the classical Liouville or Vlasov-like equation
describing the conservation of optical quasi-particles in
phase-space, cf.\ Ref.\ \cite{Akhiezer}. To illustrate the
usefulness of this approach, we consider an application to the
case of one dimensional (1D) propagation of partially incoherent
light in a nonlinear Kerr medium, and investigate the stability of
a constant amplitude plane wave against small harmonic
perturbations. It is found that, in addition to the classical
modulational instability (MI), a new linear Landau-like damping
effect arises. This damping is due to the broadening of the Wigner
spectrum, which is associated with the random phase fluctuations.
Consequently, it is found that the partially incoherent character
of the light may suppress the modulational instability, in
agreement with the result of Ref.\ \cite{Soljacic, Anastassiou}.
Finally, we illustrate an analogue to the BGK waves in plasma
physics. In the limit of the geometrical optics approximation, we
derive the Wigner distribution functions of a new class of
stationary, self-trapped, and incoherent wave pulse structures,
which may exist in a wide class of nonlinear media.

The approach proposed here sheds new light on the physics behind
the recently reported results in this field. It also allows the
formulation of conceptually new problems regarding the dynamics of
partially incoherent waves and beams in nonlinear media. In
addition, using the Wigner transform for studying this kind of
phenomena establishes new connections to quantum mechanics, plasma
physics, mesoscopic physics, image and signal processing, and
mathematics. Furthermore, the Wigner transform, apart from serving
as a very convenient mathematical tool, is widely used in its own
right in the field of optics, as a supplement to the envelope wave
function \cite{josaa}.

As our starting point, we assume that the three-dimensional (3D)
wave propagation in a dispersive (or diffractive) nonlinear medium
is described by a system of coupled model equations for the slowly
varying complex wave amplitude $\psi(t,\mathbf{r}) $ and the
nonlinear response function of the medium $n(t,\mathbf{r})$:

   \begin{equation}
   \label{e:system}
   \left\{
   \begin{aligned}
   & i\left(\frac{\partial}{\partial t}+\mathbf{v}_g\cdot\nabla\psi \right)
   \psi+\frac{\beta}{2}\nabla^2\psi+n\psi=0\\*
   & \tau_m\frac{\partial n}{\partial t}+n=\kappa
   G\left(\left\langle\psi^{\ast}\psi\right\rangle\right)
   \end{aligned}
   \right.
   \end{equation}

\noindent For convenience, we use $t$ and $\mathbf{r}$ as the evolution and
spatial dispersive variables, respectively; $\mathbf{v}_g$ is the group
velocity, $\beta$ is the diffraction or second order dispersion coefficient,
$\kappa$ is the nonlinear coefficient, and the function $
G\left(\left\langle\psi^{\ast}\psi\right\rangle\right)$ characterizes the
nonlinear properties of the medium. The bracket $\langle\ldots\rangle$ denotes
the statistical ensamble average. The relaxation time of the medium response
function, $\tau_m$, is assumed to be much longer than the characteristic time
of the statistical wave intensity fluctuations, $\tau_s$. Assuming also that $
\tau_p \gg\tau_m$, where $\tau_p$ is the characteristic time scale of the
(deterministic) wave amplitude variation, we can approximate the medium
response function as $ n\approx \kappa G \left(\left\langle\psi^\ast
\psi\right\rangle\right)$. The system \eqref{e:system} then reduces to a
generalized nonlinear Schr\"odinger equation:

   \begin{equation}
   \label{e:nlse}
   i\frac{\partial\psi}{\partial t}+\frac{\beta}{2}\nabla^2\psi+\kappa
   G\left(\left\langle\psi^\ast\psi\right\rangle\right)\psi=0,
   \end{equation}

\noindent where the coordinate system has been transformed to the reference
system moving with the group velocity $\mathbf{v}_g$, and $\mathbf{r}$ is the
reduced distance according to $ \mathbf{r}\rightarrow \mathbf{r}-\mathbf{v}_g
t$.

We stress that the independent variables in Eq. \eqref{e:nlse}
should be adopted to suit the particular situation. For instance,
to describe the 2D CW spatial optical solitons  one would use $z$
as the evolution variable and restrict the nabla operator to the
transverse dimensions, while an equation similar to Eq.
\eqref{e:nlse} with $z$ as evolution variable and
$\partial^2/\partial t^2$ instead of the nabla operator would
describe temporal 1D solitons.

Replacing $i\partial/\partial t$ by $\omega$ and $ -i \nabla$ by
$\mathbf{p}$ yields the corresponding nonlinear dispersion
relation:

   \begin{equation}
   \label{e:dispersion}
   \omega=\frac{\beta}{2} p^2-\kappa
    G\left(\left\langle\psi^\ast\psi\right\rangle\right).
   \end{equation}

We will now apply the Wigner transform method \cite{Wigner}. The
$s$-dimensional Wigner transform, including the Klimontovich statistical
average, is defined as

   \begin{widetext}
   \begin{equation}
   \label{e:Wigner}
   \rho(\mathbf{p},t,\mathbf{r})=\frac{1}{(2\pi)^s}\int_{-\infty}^{+\infty}d^s
   \bm{\xi}\,e^{i\mathbf{p} \cdot\bm{\xi}}\left\langle\psi^\ast
   \left(\mathbf{r}+\bm{\xi}/2,t\right)\psi\left(\mathbf{r}-\bm{\xi}/2,t\right)
   \right\rangle,
   \end{equation}
   \end{widetext}

\noindent which satisfies $\left\langle
\psi^\ast\left(\mathbf{r},t\right)\psi\left(\mathbf{r},t\right)\right\rangle
=\int_{-\infty}^{+\infty}d^s\mathbf{p}\,\rho\left(\mathbf{p},t,\mathbf{r}
\right)$. Applying this transform, with dimensionality $s=3$, to
Eq.\ \eqref{e:nlse}, we obtain the following Wigner-Moyal equation
for the evolution of the Wigner distribution function:

   \begin{equation}
   \label{e:wm}
   \frac{\partial\rho}{\partial t}+\beta \mathbf{p}\cdot\frac{\partial\rho}{\partial
   \mathbf{r}}+2\kappa G\left(\left\langle\left|\psi\right|^2\right\rangle\right)
   \sin\left(\frac{1}{2}\frac{\overleftarrow{\partial}}{\partial
   \mathbf{r}}\cdot\frac{\overrightarrow{\partial}}{\partial
   \mathbf{p}}\right)\rho=0.
   \end{equation}

\noindent The arrows in the sine differential operator indicate
that the derivatives act to the left and right, respectively. The
sine operator is defined by its Taylor expansion:

   \begin{equation}
   \label{e:sine}
   \sin\left(\frac{1}{2}\frac{\overleftarrow{\partial}}{\partial
   \mathbf{r}}\cdot\frac{\overrightarrow{\partial}}{\partial
   \mathbf{p}}\right)=\sum_{l=0}^{\infty}\frac{\left(-1\right)^l}{\left(2l+1\right)!2^{2l}}
   \frac{\overleftarrow{\partial}^{2l+1}}{\partial \mathbf{r}^{2l+1}}\cdot
   \frac{\overrightarrow{\partial}^{2l+1}}{\partial \mathbf{p}^{2l+1}}.
   \end{equation}

For those situations where the geometrical optics approximation is
valid, we can neglect the higher order derivatives in Eq.\
\eqref{e:sine}. This approximation is valid for
$\Delta\mathbf{p}\cdot\Delta\mathbf{r}\gg 2\pi$, where
$\Delta\mathbf{p}$ is the local width of the Wigner spectrum and
$\Delta\mathbf{r}$ is the width of the medium response function
$n(t,\mathbf{r})$. We emphasize that this is a rather strong
limitation, which can only be valid in the long-wavelength limit.
Retaining only the first term in the expansion \eqref{e:sine}, we
obtain a Vlasov-like equation:

   \begin{equation}
   \label{e:Vlasov}
   \frac{\partial\rho}{\partial t}+\beta \mathbf{p}\cdot\frac{\partial\rho}{\partial
   \mathbf{r}}+\kappa\frac{\partial G\left(\left\langle\left|\psi\right|^2
   \right\rangle\right)}{\partial \mathbf{r}}\cdot
   \frac{\partial \rho}{\partial \mathbf{p}}=0.
   \end{equation}

\noindent Equation \eqref{e:Vlasov} implies the conservation of
the number of optical quasi-particles in phase-space. Using the
Liouville theorem we recover the canonical Hamilton equations of
motion for the quasi-particle with mass $\beta$, or, equivalently,
the ray equations of the geometrical optics approximation, where
$\mathbf{r}$ and $\mathbf{p}$ are the canonical variables and
$\omega$ is the Hamilton equation defined by Eq.\
\eqref{e:dispersion}:

   \begin{equation}
   \left\{
   \begin{aligned}
   \dot{\mathbf{r}}& =\frac{\partial\omega}{\partial \mathbf{p}}=\beta
   \mathbf{p},\\*
   \dot{\mathbf{p}}& =-\frac{\partial\omega}{\partial \mathbf{r}}=\kappa\frac{\partial G
   \left(\left\langle \psi^\ast\psi\right\rangle\right)}{\partial \mathbf{r}}.
   \end{aligned}
   \right.
   \end{equation}

\noindent Note that Eq.\ \eqref{e:Vlasov} is similar to the radiation transfer
equation used in \cite{Pasmanik,Bingham,Shukla}.

As an example of the applicability of the Wigner transform
approach, we consider the MI of a 1D plane wave with a constant
amplitude in a nonlinear Kerr-like medium, for which
$G\left(\left\langle\left|\psi\right|^2\right\rangle\right)
=\left\langle\left|\psi\right|^2\right\rangle$. For the case of
\emph{coherent} light propagating in a 1D Kerr medium, it is well
known from conventional stability analysis that a perturbation of
the monochromatic stationary solution
$\psi(x,t)=\psi_0\exp\left(i\kappa\psi_0^2 t\right)$ experiences a
modulational instability when $\beta\kappa>0$ and
$K^2<4\kappa\psi_0^2/\beta$, where $K$ is the wave number of the
perturbation. The instability growth rate is given by

   \begin{equation}
   \label{e:MI_growth}
   \Omega=i\frac{\beta\kappa}{2}\left(\frac{4\kappa\psi_0^2}{\beta
   K^2}-1\right)^{1/2}.
   \end{equation}

To investigate the effect of the incoherence, we assume a Wigner
distribution function of the form $
\rho(p,t,x)=\rho_0(p)+\rho_1\exp\left[i\left(K x-\Omega
t\right)\right]$, with $\rho_0\gg\left|\rho_1\right|$. Here,
$\rho_0(p)$ is the background distribution function corresponding
to the plane wave with a complex amplitude
 $\psi=\psi_0\exp\left[i\kappa\psi_0^2 t+i\phi(x)\right]$, implying
$\int_{-\infty}^{+\infty} \rho_0(p)\,dp=\psi_0^2$. The incoherence
is modeled by the randomly varying phase term $\phi(x)$. Studying
the linear evolution of the perturbation $\rho_1$, we obtain from
the linearized Wigner-Moyal equation \eqref{e:wm} the following
dispersion relation:

   \begin{equation}
   \label{e:wm_disp}
   1+\frac{\kappa}{\beta}\int_{-\infty}^{+\infty}\frac{\rho_0
   \left(p+K/2\right)-\rho_0\left(p-K/2\right)}
   {K\left(p-\Omega/\beta K\right)}\,dp=0.
   \end{equation}

\noindent The corresponding dispersion relation following from the linearized
Vlasov-like equation \eqref{e:Vlasov} has the form

   \begin{equation}
   \label{e:vlasov_disp}
   1+\frac{\kappa}{\beta}\int_{-\infty}^{+\infty}
   \frac{d\rho_0/dp}{\left(p-\Omega/\beta K\right)}\,dp=0,
   \end{equation}

\noindent which can also be directly obtained from Eq.\
\eqref{e:wm_disp} by taking the limit of small $K$. The relation
\eqref{e:vlasov_disp} is similar to the dispersion relation for
electron plasma waves, which is well known to contain the effect
of the Landau damping. In general, the kinetic integrals in
\eqref{e:wm_disp} and \eqref{e:vlasov_disp} can be represented as
the sum of a principal value and a residue contribution, where the
latter leads to a Landau-like damping of the perturbation. This
stabilizing effect is not an ordinary, dissipative damping. It is
rather an energy-conserving self-action effect within a partially
incoherent wave field, which causes a redistribution of the Wigner
spectrum because of the interaction between different parts of the
spectrum. This spectral redistribution counteracts the MI. Similar
phenomena occur in connection with nonlinear propagation of
electron plasma waves interacting with intense electromagnetic
radiation \cite{Venedov,Bingham}, nonlinear interaction between
random phase photons and sound waves in electron-positron plasmas
\cite{Shukla}, and the longitudinal dynamics of charged-particle
beams in accelerators \cite{Anderson}.

It is interesting to note that for a coherent wave, i.e., a delta-shaped
background distribution $\rho_0(p)=\psi_0^2\delta(p)$, and for $\beta\kappa>0$,
Eq.\ \eqref{e:wm_disp} gives exactly the modulational instability growth rate
defined by Eq.\ \eqref{e:MI_growth}. For the same case, the dispersion relation
\eqref{e:vlasov_disp} yields $\Omega=i\left(\beta\kappa\right)^{1/2}\psi_0 K$,
which is identical to the growth rate of the MI as given by the expression
\eqref{e:MI_growth} in the limit of long wavelengths, i.e., when
$K^2\ll4\kappa\psi_0^2/\beta$.

To illustrate the incoherent case, we assume that $\phi(x)$ is described by the
following autocorrelation function:

   \begin{equation}
   \label{e:rphi}
   \left\langle\exp\left[-i\phi\left(x+y/2\right)+i\phi\left(x-y/2\right)\right]\right\rangle=
   \exp\left(-p_0\left|y\right|\right),
   \end{equation}

\noindent where $p_0^{-1}$ is the correlation length. The corresponding Wigner
function has a Lorentzian shape:

   \begin{equation}
   \label{e:lorentz}
   \rho_0(p)=\frac{\psi_0^2}{\pi}\frac{p_0}{p^2+p_0^2},
   \end{equation}

\noindent and the dispersion relation \eqref{e:wm_disp} yields the following
exact result for $4\kappa\psi_0^2/\beta K^2>1$:

   \begin{equation}
   \label{e:landaudamp}
   \frac{\Omega}{\beta K}=i\frac{K}{2}
   \left(\frac{4\kappa\psi_0^2}{\beta K^2}-1\right)^{1/2}-ip_0,
   \end{equation}

\noindent which is similar to the result obtained in Ref.\ \cite{Soljacic}.

Equation \eqref{e:landaudamp} clearly shows the stabilizing effect of the
Landau-like damping due to the finite width $p_0$ of the Lorentzian spectrum
(or the finite correlation length of the wave phase). In fact, if the width of
the Lorentzian spectrum, $p_0$, satisfies the relation $p_0> p_c\equiv K_c/2$,
where $K_c\equiv \left(4\kappa\psi_0^2/\beta\right)^{1/2}$ is the cut-off
wavelength of the MI, the instability is completely suppressed for all wave
numbers $K$ of the perturbation. In other words, if the Landau-like damping,
induced by the broadening of the Wigner spectrum due to the partial incoherence
of the wave, is strong enough, it can overcome the coherent growth associated
with the MI. This result has also been verified experimentally; see, e.g.,
Ref.\ \cite{Anastassiou}.

Finally, we will demonstrate the existence of a class of
self-trapped and partially incoherent wave pulse solutions to the
stationary Vlasov-like equation

   \begin{equation}
   \label{e:bgk}
   \beta p\frac{\partial\rho}{\partial x}+\kappa
   \frac{\partial G\left(\left\langle\left|\psi\right|^2\right\rangle\right)}{\partial x}
   \frac{\partial \rho}{\partial p}=0
   \end{equation}

\noindent for an arbitrary nonlinear function
$G\left(\left\langle\left|\psi\right|^2\right\rangle\right)$.
According to the Jeans theorem, c.f.\ \cite{Akhiezer}, the
solution of Eq.\ \eqref{e:bgk} can be expressed as an arbitrary
function of the Hamiltonian $H=\beta p^2/2-\kappa
G\left(\left\langle\left|\psi\right|^2\right\rangle\right)$. Thus,
we have $\rho=\rho_s(H)$, where $\rho_s$ is an arbitrary function.
The quasi-particles are trapped in the nonlinear potential for
$-\gamma\leq p\leq\gamma$, where $\gamma=\left(2\kappa
G\left(\left\langle\left|\psi\right|^2\right\rangle\right)/\beta\right)^{1/2}$
and $\kappa\beta>0$, and consequently the condition
$\int_{-\gamma}^{+\gamma}dp\,\rho_0(p)=\left\langle\left|\psi\right|^2\right\rangle$
leads to an integral equation having the following solution for
$\rho_s$ (c.f.\ \cite{Bernstein}):

   \begin{equation}
   \label{e:bernstein}
   \rho_s(H)=\frac{1}{\pi}\left(\frac{\beta}{2}\right)^{1/2}
   \int_{0}^{-H}\frac{dH'}{\left(-H-H'\right)^{1/2}}
   \frac{dF\left(H'\right)}{dH'},
   \end{equation}

\noindent where $F\left(\Theta\right)\equiv G^{-1}\left(\Theta\right)$ with
$\Theta\equiv
G\left(\left\langle\left|\psi\right|^2\right\rangle\right)=\left(\beta
p^2/2-H\right)/\kappa$. For instance, for the Kerr nonlinearity we have
$F\left(\Theta\right)\equiv\Theta$ and Eq.\ \eqref{e:bernstein} yields
$\rho_s(H)=\left(2/\pi\kappa\right)\sqrt{-\beta H/2}$. Equation
\eqref{e:bernstein} describes the Wigner distributions of quasi-particles
trapped in a collectively produced, stationary, partially incoherent, and
localized wave structure. These structures are similar to the large-amplitude
BGK waves in a plasma; see \cite{Bernstein}.

In conclusion, we have proposed a general, statistical approach
for the theoretical analysis of the propagation of partially
incoherent optical waves and beams in dispersive media with
arbitrary nonlinearities. The approach is based on the Wigner
transform method including the Klimontovich statistical average.
The derived Wigner-Moyal equation determining the evolution of the
Wigner distribution function represents a generalization of the
Vlasov-like equation, which is only valid within the geometrical
optics approximation. The Wigner-Moyal equation clearly shows that
the number of optical quasi-particles is \emph{not} conserved in
phase-space beyond the validity of the geometrical optics
approximation. Using the Wigner-Moyal equation, we have carried
out a linear stability analysis for small perturbations on a
constant, 1D, partially incoherent background in a nonlinear Kerr
medium. The theory reproduces the exact expression for the MI
growth rate of a coherent wave, but also includes a linear
Landau-like damping effect associated with the broadening of the
Wigner spectrum due to partial wave incoherence. This damping
effect explains the previously reported incoherent suppression of
the modulational instability.

\begin{acknowledgments}
The authors are grateful to D.\ N.\ Christodoulides, M.\ Segev and A.\ Hasegawa
for valuable discussions.
\end{acknowledgments}


%
%

\end{document}